\documentclass[preprint,tightenlines]{revtex4}
\usepackage{graphics}

\newcommand{\eref}[1]{Eq. (\ref{#1})}
\newcommand{\fref}[1]{Fig. (\ref{#1})}
\newcommand{\mycite}[1]{Ref. \cite{#1}}
\newcommand{\beq}{\begin{equation}}
\newcommand{\eeq}{\end{equation}}
\newcommand{\bea}{\begin{eqnarray}}
\newcommand{\eea}{\end{eqnarray}}
\newcommand{\sstw}{\ensuremath{{\rm sin}^2\theta_W}}
\newcommand{\enumax}{\ensuremath{E_\nu^{\rm max}}}
\newcommand{\costh}{\ensuremath{{\rm cos}\,\theta}}
\newcommand{\bigc}{\ensuremath{\mathcal{C}}}

\begin{document}

\title{Manipulating a Neutrino Spectrum to Maximize the Physics Potential from a Low Energy Beta Beam}

\author{Philip S. Amanik and Gail C. McLaughlin}
\affiliation{Department of Physics, North Carolina State University, Raleigh, NC 27695}

\date{\today}

\begin{abstract}
Proposed low energy beta beam facilities would be capable of producing intense beams of neutrinos (anti-neutrinos) with well defined spectra. We present analytic expressions and numerical results which accurately show how the total neutrino flux reaching the detector depends on the geometry of the source and the detector. Several authors have proposed measurements which require using different flux shapes. We show that detectors of different sizes and shapes will receive neutrino fluxes with different spectral shapes, and that the spectral shape will also be different in different regions of the same detector. Our findings also show that for certain measurements systematic uncertainties and run time can be reduced. 
\end{abstract} 


\maketitle

\section{Introduction}
The concept of a beta beam facility to produce neutrinos and anti-neutrinos has recently been proposed \cite{zucchelli} and feasibility studies have been performed \cite{autin} and are underway \cite{bbweb}. The source of the neutrinos will be radioactive ions which undergo beta decay.  The ions will be boosted to relativistic speeds and contained in a storage ring which has two straight sections. Neutrinos emitted along a straight section will reach a detector placed along the axis of the straight section.  The neutrino energy spectrum in the ion rest frame is known theoretically and verified by experiment.  The spectrum from the boosted ions can thus be calculated and is well defined.  The beta beam neutrino source is appealing because it will produce neutrinos of a single flavor\cite{zucchelli}.  Experiments to measure neutrino oscillation parameters, including the CP violating phase, where proposed in Refs \cite{zucchelli,mezzetto}. 

A low energy beta beam was considered in \cite{CV1} (for a recent review see Ref. \cite{CV2}) to produce neutrinos with energy in the range of 10 to 100 MeV. This energy range is relevant for supernova neutrinos, and low energy neutrino physics in general.  Typical boost factors that would generate neutrinos in this range are $5 \leq \gamma \leq 15$. Different sets of ring dimensions for a low energy beta beam, the resultant neutrino fluxes, and neutrino-nucleus inelastic scattering event rates at a close detector ($\sim$ 10 meters from the ring) were studied in detail in Ref. \cite{S&V}. There it was found that the smallest possible ring, with as large as possible straight section for that ring, would result in a higher number of neutrinos reaching the target. This conclusion was independently confirmed in \cite{mclaughlin}.  

Several experiments to be done at a low energy beta beam have been proposed.  These include measuring neutrino-nucleus inelastic scattering cross sections\cite{S&V}, testing the response of a lead based supernova neutrino detector\cite{mclaughlin}, detecting a neutrino magnetic moment by measuring neutrino electron scattering\cite{M&V}, measuring \sstw 
with $\nu_e-e^-$ scattering\cite{BJV}, testing conservation of the vector current by measuring $\bar{\nu}_e-p$ 
capture\cite{BJLV}, reconstructing a detected supernova spectrum as a linear combination of beta-beam neutrino 
spectrum\cite{J&M} and measuring \sstw with neutrino-nucleus elastic scattering\cite{bueno}. A unique feature of the beta beam neutrino source is that it is possible to produce different spectra shapes with different energy ranges by boosting the parent ions to different $\gamma$ factors. This feature was first exploited in \mycite{mclaughlin} and then later in References \cite{BJV,J&M} for the measurements they propose. References \cite{S&V,bueno} also suggest using this feature to measure cross sections at different energies. 

In this paper we derive, for a special case, an exact analytical expression that gives, for any point along the straight section, the percentage of the total emitted neutrinos that are contained in a lab angle $\theta_L$.  This expression is useful for considering detector sizes that will both receive a desired flux and maximize the number of neutrinos received. It is also useful for understanding the shape of the spectra at the detector.  Using our expression, we give examples that show what percentage of neutrinos, from different points on the ring, reach the detector and what percentage of the neutrinos, from the same point on the ring, reach different parts of the detector.  

We consider for the first time the flux reaching detectors of various sizes. In particular, we calculate the neutrino flux reaching cylindrical detectors of the same mass but with different dimensions. We show that changing the radius and length of a cylindrical detector will alter the total flux and spectra shape reaching the detector.  We also show that inner and outer regions of a cylindrical detector will receive different flux and spectra shape. Given this finding we propose that instead of running the beam at different energies in order to produce a different shaped flux in the detector, the beam could be run at a fixed energy and events can be considered from different regions of the detector (that receive different flux shapes). This method to produce different shaped neutrino fluxes in a beta beam experiment has the advantage that no new systematic uncertainties are introduced from running the beam at different energies.  This method could therefore allow for more accurate measurements and simplify fitting analysis.  

We calculate expected neutrino scattering events that would occur in the inner and outer regions of a cylindrical detector and compare these to events occurring from running the beam with ions boosted to different $\gamma$ factors. The event rates show that for certain measurements, more events can be obtained by using our method for producing different flux shapes in a detector. Thus our proposed method for producing different shaped neutrino fluxes not only can reduce systematic error, it can also reduce the beam time needed.

The average energy of neutrinos emitted from a beam of relativistic particles will have some dependence on the angle of emission relative to the beam axis.  This is a consequence of special relativity, and we take advantage of it in our considerations of the neutrino fluxes reaching different regions of an on-axis detector.  This property can also be exploited for use in another context---putting a detector off a beam axis, e.g. for pion decay in flight \cite{t2k,nova} and recently for a beta beam \cite{LBJV}.

The paper is organized as follows. In section II we derive the expression for the percentage of neutrinos emitted from a point on the straight section that are contained in a given lab angle. We also present figures to demonstrate the use of this expression. In section III we discuss different detector shapes and present plots of the calculated flux reaching these detectors. In section IV we examine the flux reaching different regions of a detector and we calculate neutrino scattering event rates in these regions. We show how certain experiments can be made more efficient by exploiting our findings regarding the flux reaching seperate regions of the detector. We give conclusions in section V. Our formalism for calculating the neutrino flux in the lab frame is presented in the appendix. Throughout the paper we use units in which 
$\hbar=c=1$.

\section{Number of Neutrinos Emitted in a Lab Angle}
A currently used estimate for the amount of emitted flux in the lab frame is that most of the flux is contained in an angle of $1/\gamma$, where $\gamma$ is the boost factor of the ions. For example, this estimate was used in Ref. \cite{S&V}. 
(Note that Ref. \cite{S&V} also provided exact calculations of numbers of events.) The estimate may be used to choose detector size parameters suitable for capturing most of the flux coming from the straight section.  However, one may desire a more accurate estimate. In particular, we will show that maximizing the number of neutrinos that reach the detector will not necessarily maximize the flux.

We aim to quantify how many neutrinos are emitted in a lab angle. More specifically, given any point along the straight section of the storage ring, what fraction of the total neutrinos emitted from that point at any instant in time are contained in a given lab angle? In the appendix, we present our formalism which expresses the probability that a neutrino will be emitted with a certain energy and angle in the lab frame. To answer our present question we refer to \eref{restspect} and the coordinate transformation of Eq.s (\ref{elab}) - (\ref{plabt}) and consider a special case. Instead of using the rest frame beta decay spectrum of \eref{restspect}, we will use a monoenergetic rest frame spectrum defined by a delta function
\beq
f_R(E_R) = \bigc\,\delta (E_R - E_0), \label{deltaspec}.
\eeq
Here, \bigc \ has dimensions of per second and the delta function has implicit dimensions of per energy. Using \eref{theta-flux} and transforming this flux in the manner of \eref{trnsfrmtn} we have
\beq
1 = \int^1_{-1}\int^{E_L^{\rm max}}_0 \frac{\bigc\,\delta[E_L\gamma(1-v\costh_L) - E_0]}{2\,\bigc}
			\frac{dE_L\,d\costh_L}{\gamma(1-v \costh_L)}, \label{deltatrans}
\eeq
where $v$ is the relativistic speed of the nuclei and $\gamma = 1/\sqrt{1-v^2}$ is the boost factor. We know from the discussion in the appendix Sec. \ref{appendix-transformation} that if we restrict the bounds of integration in \eref{deltatrans}, then \eref{deltatrans} will give normalized probability that a neutrino will be emitted with lab energy $\in (E_L^1, E_L^2)$ and lab direction cosine $\in (\costh_L^1, \costh_L^2)$.  For the case of a monoenergetic spectrum there is only one energy represented in the rest frame.  However, since the rest frame angle is arbitrary, that one energy can be transformed to a range of energies in the lab frame. Equation 
(\ref{deltatrans}) is useful since the integrations can be performed analytically using the identity
\beq
\delta \Big( f(y) \Big ) = \sum_i \left | \frac{df}{dy}\Bigg |_{y_i} \right |^{-1}\delta(y-y_i),
\eeq
where $y_i$ are simple zeros of $f(y)$.

The upper and lower bounds on the \costh \ integration in \eref{deltatrans} correspond to angles $\theta = 0$ and 
$\theta = \pi$, respectively.  We may change the lower bound to $\costh^{\rm open}$ corresponding to some opening angle 
$\theta^{\rm open}$. In this case, performing the integrations in \eref{deltatrans} would give the probability that a neutrino is emitted in the lab frame between $\theta^{\rm open}$ and $\theta = 0$.  Denoting this probability 
${\rm P}^{\rm open}$, we have
\beq
{\rm P}^{\rm open} = \frac{1- \costh^{\rm open}}{2\gamma^2(1-v)(1-v\,\costh^{\rm open})}. \label{Popen}
\eeq
Notice that this result depends only on the speed $v$ to which the ions are boosted. This makes sense given \eref{thetaL} which shows that the lab frame angle only depends on $v$ and $\costh_R$, the angle the neutrino was emitted from in the rest frame.  If the neutrino is emitted in a more forward direction in the rest frame ($\costh_R$ closer to 1), \eref{elab} implies the neutrino will have a higher energy in the lab frame. Thus, the amount of neutrinos that get forward boosted within a given lab angle depends on the speed of the nuclei, and neutrinos with smaller lab angles will have a higher lab energy than do neutrinos with larger lab angles. These considerations apply to our current case of a monoenergetic rest frame spectrum, but they generalize to the case of the distribution spectrum given by \eref{restspect}. In the next section we will see the consequences, for the distribution spectrum, of the fact that more forward boosted neutrinos have higher energies than do neutrinos with large lab angles.
 
In the appendix Sec. \ref{appendix-integrating} we explain that $\omega/L$ (where $\omega$ is the number of nuclei/sec injected in the storage ring and $L$ is the total length of the ring) is the number of neutrinos emitted per second from a line element $dx$ on the ring. Multiplying ${\rm P}^{\rm open}$ by $\omega/L$ gives, for any point, the number of neutrinos per second emitted inside angle 
$\theta^{\rm open}$, where the angle is measured from the point of emmission.  We now define $\xi^{\rm open}$, the fraction of neutrinos emitted from some point on the straight section that are contained in angle $\theta^{\rm open}$. Since the total number of neutrinos per second emitted from any point on the straight section is $\omega/L$ we have that 
\beq
\xi^{\rm open} = \frac{\frac \omega L {\rm P}^{\rm open}}{\omega/L}. 
\eeq
So we see that $\xi^{\rm open} = {\rm P}^{\rm open}$. We can now calculate, for the case of a monoenergetic rest frame spectrum, exactly what fraction of the neutrinos emitted from any point on the straight section are contained in a lab angle $\theta_L\sim 1/\gamma$ for a beam run at 
$\gamma = 7$. Using $\xi^{\rm open}$ we find that only $50\%$ of the neutrinos emitted at each point are contained in this angle for this case.

\begin{figure}
\scalebox{0.5}{\includegraphics{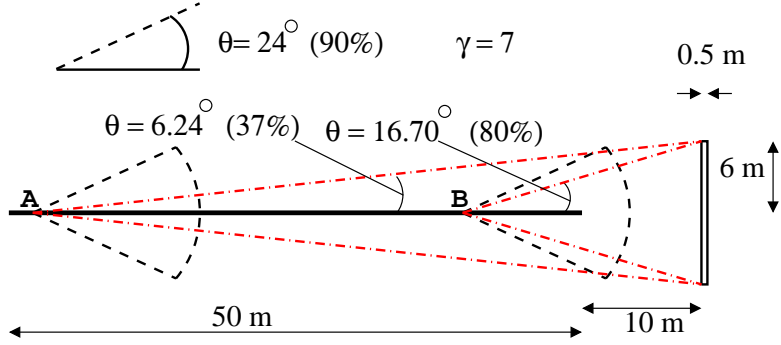}}
\caption{(Color Online) Example configuration to show the use of $\xi^{\rm open}$ to calculate the fraction of neutrinos emitted from two arbitrary points along the straight section, that reach the detector. This example is for nuclei boosted to $\gamma =7$. The figure is to scale. The thick black line represents the straight section, the rectangle is a side view of the cylindrical detector. The dashed cones show the lab angle that contains $90 \%$ of the emitted neutrinos.  The dashed-dot lines show the opening angle to the detector face.}
\label{fig:2points}
\end{figure}

\begin{figure}
\scalebox{0.5}{\includegraphics{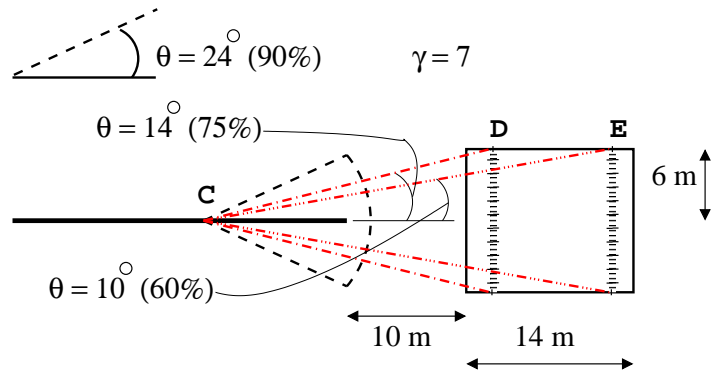}}
\caption{(Color Online) Example configuration to show the use of $\xi^{\rm open}$ to calculate the fraction of neutrinos emitted from an arbitrary point along the straight section, that reach two arbitrary slices of the detector. This example is for nuclei boosted to $\gamma =7$. The figure is to scale. The thick black line represents the straight section, the rectangle is a side view of the cylindrical detector. The dashed cones show the lab angle that contains $90 \%$ 
of the emitted neutrinos.  The dashed-dot lines show the opening angle to the two detector slices.}
\label{fig:2slices}
\end{figure}

To illustrate the use of $\xi^{\rm open}$, we consider two examples of detector configurations.  The dimensions of the detectors in these examples are not intended to suggest an actual experimental setup, but rather to demonstrate the use of 
$\xi^{\rm open}$.  In the first example we show the fraction of neutrinos emitted from two different points on the straight section that reach the same region of the detector. In the second example we show the fraction of neutrinos emitted from the same point on the straight section that reach two different regions of the detector. These examples are shown in \fref{fig:2points} and \fref{fig:2slices}, which are to scale. In each figure, the thick black line is the straight section of the storage ring, the rectangle is a side view of the cylindrical detector, and the dashed cone shows the opening angle that contains $90\%$ of the neutrinos emitted from that point. 

In \fref{fig:2points}, we show a detector with radius 6m and thickness 0.5m.  The detector is 10m away from the straight section, which itself is 50m. The opening angle from point B to the detector is $16.7^\circ$ which, using 
$\xi^{\rm open}$, contains $80\%$ of the neutrinos emitted from that point. Thus, only $80\%$ of the neutrinos emitted from that point reach the detector.  The opening angle from point A to the detector is $6.24^\circ$, which contains 
$37\%$ of the neutrinos emitted from that point. 

In \fref{fig:2slices}, we show a detector with radius 6m and thickness 14m that is located 10m away from the straight section.  The opening angle from point C on the straight section to the detector slice at D is $14^\circ$. Using $\xi^{\rm open}$, we calculate that $75\%$ of the neutrinos emitted from point C reach this slice of the detector.
The angle from point C to the detector slice at E is $10^\circ$ and $60\%$ of the neutrinos emitted from point C reach this slice. This figure also shows that the neutrinos emitted between opening angle of $14^\circ$ and $10^\circ$ only have a small portion of the detector to interact with.

\section{Modifying Detector Dimensions}
\subsection{Number of Neutrinos vs. Flux}
In the previous section we saw how to quantify the number of neutrinos reaching the detector. We now study how changing detector dimensions will effect the neutrino flux in the detector. It is apparent from \fref{fig:2points} that a larger radius will allow the detector to catch more neutrinos. However, the flux is the number of neutrinos (per second per energy) reaching a slice of the detector \emph{divided} by the area of that slice. Increasing the radius has two effects. Allowing more neutrinos to reach the detector works towards increasing the flux. Dividing by a larger area will reduce the average flux.  

It is apparent from \fref{fig:2slices} that if the thickness of the detector is increased less neutrinos will reach the back end of the detector than will reach the front end of the detector. However, the neutrinos that do reach the back of the detector will have had opportunity to interact with more detector material than do neutrinos emitted with larger lab angles. The neutrinos that are exposed to more detector material will have a greater chance of having an interaction within the detector. 

In the case of having a fixed thickness, increasing the radius of the detector will increase its total mass. Likewise, for a fixed radius, increasing the thickness will also increase the total mass of the detector. Larger mass detectors of course have more targets and the potential for more events. 

\begin{figure}
\scalebox{0.5}{\includegraphics{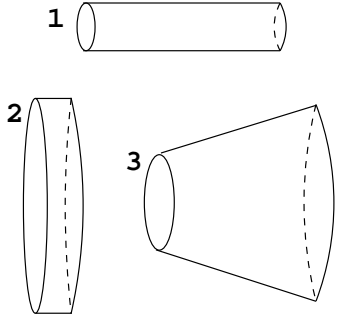}}
\caption{Examples of shapes for a detector. }
\label{fig:shapes}
\end{figure}

Motivated by Figures (\ref{fig:2points})-(\ref{fig:2slices}) and these qualitative considerations, we use \eref{flux} to calculate the average flux in detectors of different radii and thickness. We would like to study whether a long detector with small radius (see detector 1 in \fref{fig:shapes}) would be more desirable than a short detector with a large radius (detector 2 in \fref{fig:shapes}).  When comparing detectors with different radii and thickness it will be useful to have some common property of the detectors. Therefore, we will keep the volume fixed and vary the radius and thickness while maintaining the volume.  For our example, we chose a detector with volume $71.2\,{\rm m}^3$ and we use five different pairs of radius $r$ and thickness $h$, both in meters.  The pairs we have chosen, denoted as $(r,h)$, are (2.13, 5), (2.75, 3), (3.89, 1.5), (4.77, 1) and (6.74, 0.5).  We calculate the fluxes for the case of a beta beam facility with storage ring of total length $L = 450\,{\rm m}$, straight section length $S = 150\,{\rm m}$, detector located $d=10\,{\rm m}$ away from the straight section and with $\omega = 5\times 10^{11}$ nuclei/sec injected into the ring \cite{S&V}. We do our calculations for a beam composed of $^{18}{\rm Ne}$ nuclei. $^{18}{\rm Ne}$ has a half life of 1.67 seconds and $\beta^+$ decays emitting a neutrino with a maximum possible energy of 3.42 MeV.  We use this configuration of the beam for all our calculations throughout the paper. The fluxes reaching the detectors mentioned above are plotted in \fref{fig:chngdims7-14} for the cases of a beam run at boost of $\gamma=7$ and $\gamma=14$.  For comparison, in \fref{fig:chnggams} we show different flux plots calculated for a beam run at boost factors of $\gamma=7,8,9,12,14$ with a $71.2\,{\rm m}^3$ detector with r=2.13m and h=5m.

\subsection{Flux Curves} \label{sec:altflux}
\begin{figure}
\scalebox{1.2}{\includegraphics{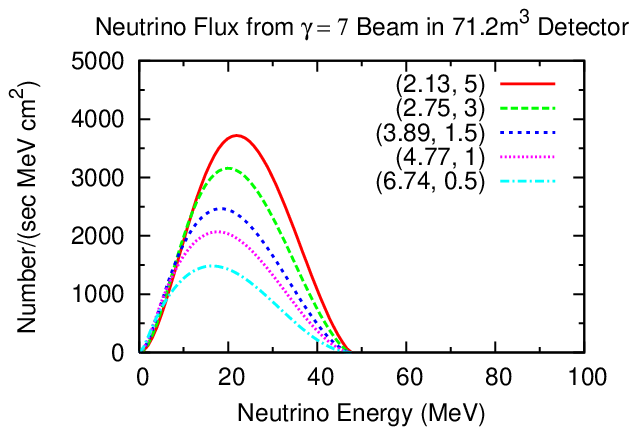}}
\scalebox{1.2}{\includegraphics{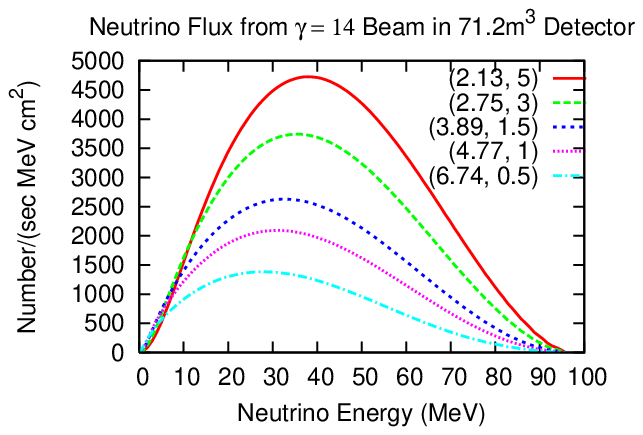}}
\caption{(Color Online) Neutrino fluxes in detector with volume $71.2\,{\rm m}^3$ for five pairs of cylinder radius $r$ and thickness $h$, both in meters. Pairs are denoted as $(r,h)$ in the key.  Fluxes were calculated for storage ring of total length $L = 450\,{\rm m}$, straight section length $S = 150\,{\rm m}$, detector located at $d = 10\,{\rm m}$ away from the straight section and $\omega = 5\times10^{11}$ nuclei/sec injected in the ring.  In the left panel, fluxes are calculated for a beam run at a boost of $\gamma = 7$; in the right panel, fluxes are calculated for a beam run at a boost of $\gamma = 14$.}
\label{fig:chngdims7-14}
\end{figure}

\begin{figure}
\scalebox{1.2}{\includegraphics{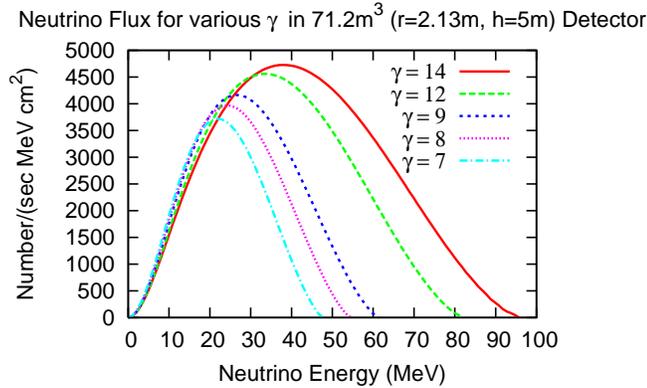}}
\caption{(Color Online) Neutrino fluxes, for beam run at boost factors of $\gamma=7,8,9,12,14$, in detector a $71.2\,{\rm m}^3$ detector with r=2.13m and h=5m. Experimental setup assumed is the same as that for the fluxes calculated in \fref{fig:chngdims7-14}.}
\label{fig:chnggams}
\end{figure}

The plots in \fref{fig:chnggams} demonstrate the attractive feature of the beta-beam neutrino experiment, that it is capable of producing neutrino fluxes of different shapes by boosting the nuclei to different $\gamma$ factors. This feature was utilized for the measurements proposed in References \cite{mclaughlin,BJV,J&M}. References \cite{S&V,bueno} also mentioned taking advantage of this feature. The plots in \fref{fig:chngdims7-14} illustrate our conclusion that \emph{another} way to produce different shaped neutrino fluxes is by using different shaped detectors with a beam run at only one boost factor.

From \fref{fig:chngdims7-14} we see that for the detector dimensions we have chosen in this example, the flux decreases as the radius of the detector increases.  This is consistent with our qualitative observations above. When considering detector sizes, beyond technical and engineering restrictions, one should strive to find a balance between maximizing the number of neutrinos reaching the detector and maximizing the flux within the detector.

We also notice from \fref{fig:chngdims7-14} that as the radius decreases the average energy for the flux distributions decreases. This is consistent with what we found in the previous section for the case of a monoenergetic rest frame spectrum - that neutrinos emitted with smaller angles in the lab frame, i.e. more forward boosted, have higher energies. Obviously, the distribution rest frame spectrum is more complicated and does not behave exactly as a mono-energetic spectrum does when transformed to the lab frame.  However, \fref{fig:chngdims7-14} clearly shows that there are more higher energy neutrinos in the forward boosted direction.  

We see from \fref{fig:chngdims7-14} that if we produce different flux distributions in the detector by changing the size of the detector, we will always have neutrinos with energy in the same range [0, $E_L^{\rm max}$]. Comparing to \fref{fig:chnggams}, where different flux distributions are produced by boosting the nuclei to different $\gamma$ factors, we see that the available energy range of the neutrinos decreases as $\gamma$ decreases. Thus, changing the spectrum by changing the detector could be useful if we want to compare different measurements involving higher energy neutrinos.   

Another potential advantage of changing the flux shape by changing detector dimensions is that no new beam systematic uncertainties are introduced. If the beam is run at different boost factors there may be different losses during acceleration, or different losses due to nuclei decaying before reaching the ring. However, if uncertainties stemming from using different sized detectors are lower than beam uncertainties, it would be favorable to produce different shaped spectra by running the beam at a fixed energy and using different sized detectors.
 
\section{Flux Shapes and Event Rates in Different Regions of Detector}

\subsection{Flux Shape in Seperate Regions of Detector}
\begin{figure}
\scalebox{0.5}{\includegraphics{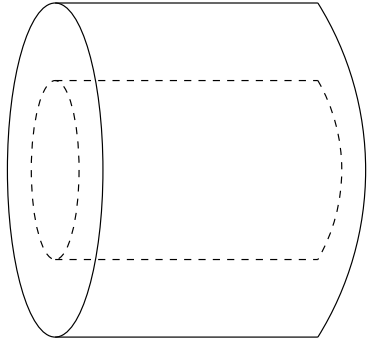}}
\caption{Cylinder detector separated into an inner cylinder and an outer hollow cylinder region. }
\label{fig:regions}
\end{figure}

\begin{figure}
\scalebox{1.2}{\includegraphics{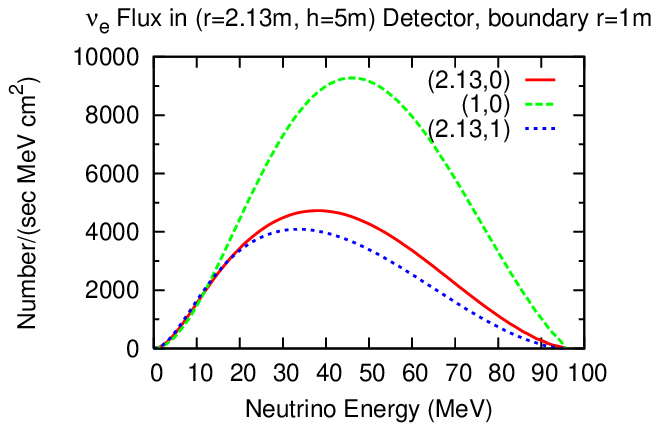}}
\scalebox{1.2}{\includegraphics{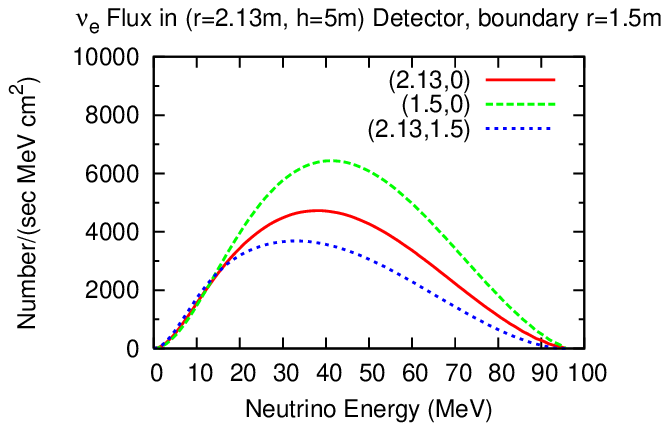}}
\caption{(Color Online) Neutrino fluxes in detector with volume $71.2\,{\rm m}^3$ and r=2.13m and h=5m. Fluxes were calculated for separate inner and outer regions of the detector. Pairs in the key denote the radii of the cylinder regions - the first number is the location of the outer radius and the second number is the location of the inner radius. Fluxes were calculated for a storage ring of total length $L = 450\,{\rm m}$, straight section length $S = 150\,{\rm m}$, detector located at $d = 10\,{\rm m}$ away from the straight section, $\omega = 5\times10^{11}$ nuclei/sec injected in the ring and a beam run at a boost of $\gamma = 14$. The solid curve shows the flux in the whole region, the long dash curve shows the flux in the inner region and the short dash curve shows the flux in the outer region. For the plots on the left panel, the boundary between the inner and outer regions is located at r=1m; for the plots in the right panel the boundary is at r=1.5m.}
\label{fig:fluxcuts}
\end{figure}

We now present another new way for obtaining different flux shapes in a beta beam experiment. Building on the results of the previous two sections, we investigate the shape of the flux in different regions of one detector.  We divide a cylindrical detector into inner cylinder and outer hollow cylinder regions, as in \fref{fig:regions}, and find that different flux shapes reach these seperate regions.  This may not come as a surprise given the results shown in the previous two sections. However, to illustrate this result we present plots of calculated fluxes for the inner and outer regions of a cylinder detector.  Continuing with our current example, we use a $71.2\,{\rm m}^3$ detector with r=2.13m and h=5m. We divide this detector into seperate inner cylinder and outer hollow cylinder regions.  We consider two cases for separation: one with the boundary at radius r=1m and another with the boundary at radius r=1.5m. For the case with boundary at r=1m, the inner cylinder has volume $15.7\,{\rm m}^3$ and the outer hollow cylinder region has volume $55.6\,{\rm m}^3$. For the case with the boundary at r=1.5m, each region has a volume of about $35.5\,{\rm m}^3$.  In each case, we use the same parameters for the experimental setup that we have been using: storage ring of total length $L = 450\,{\rm m}$, straight section length $S = 150\,{\rm m}$, detector located $d=10\,{\rm m}$ away from the straight section and $\omega = 5\times 10^{11}$ nuclei/sec injected into the ring. We show plots of the calculated fluxes for these two cases, for nuclei boosted to $\gamma=14$, in 
\fref{fig:fluxcuts}.

We can see from \fref{fig:fluxcuts} that the neutrino flux in the inner and outer regions of a detector do have different shapes. In each figure we see that the flux in the inner region has a higher peak than the flux over the whole detector. Thus we see that there is a greater concentration of the neutrinos in the inner region of the detector. We see also that the average energy of neutrinos reaching the inner region is greater than the average energy of neutrinos reaching the outer region.  We therefore conclude that another way to achieve different shaped fluxes in a beta beam experiment is to separately count events that occurred in an inner region and outer region of a detector.  Location for events can be identified in Cherenkov and scintillator detectors.  Instead of running the beam for one year at some boost factor and then another year at a different boost factor, or running the beam for one year with one detector and another year with a different sized detector, the beam could be run for two years with one detector and events could be taken from two seperate regions.  This method has the advantage that no new systematic uncertainties are introduced from using different detectors or different beams.

\subsection{Events in Seperate Regions of Detector}
We illustrate this method for obtaining different flux shapes by calculating event rates for neutrino-nucleus elastic scattering.  This process was studied in \mycite{bueno} at a beta beam facility and also in \mycite{scholberg} for neutrinos produced at a stopped-pion source. For our example we consider a liquid noble nuclear recoil detector based on CLEAN\cite{clean}.  This detector concept has been proposed with a spherical geometry, but cylindrical geometries would also be possible \cite{clean}. We calculate events for a 100 tonne fiducial volume in a cylindrical detector containing liquid argon. Liquid argon has a density of 
$1.4\,{\rm g}/{\rm cm}^3$ so 100 tonnes will occupy a volume of $71\,{\rm m}^3$. For our example we will choose a fiducial region with radius of 2.13m and thickness of 5m. An actual detector would have a larger total mass and larger total dimensions and the region surrounding the fiducial volume would be used to reject some background events from external sources. Since the purpose of our example here is to illustrate how different flux shapes could be achieved by separating events from different regions of a detector, and not intended for proposing a specific measurement, we do not address the issue of backgrounds that would exist in the detector we are considering.  Our finding is general and applies not only to liquid noble detectors, but any detector that has position resolution for neutrino interaction events.  

This method applies to detectors of other shapes as well. For example, recalling from Section \ref{sec:altflux} that forward boosted neutrinos have higher energy, a cone shaped detector (such as detector 3 in \fref{fig:shapes}) would be useful for selecting higher energy neutrinos. Engineering and technical limitations may prevent a realistic detector of this shape but such a detector could be simulated in a cylindrical detector. By selecting events that occur within a cone shaped region of a detector we could pick out events which involved mainly higher energy neutrinos.
 
\begin{figure}
\scalebox{1.2}{\includegraphics{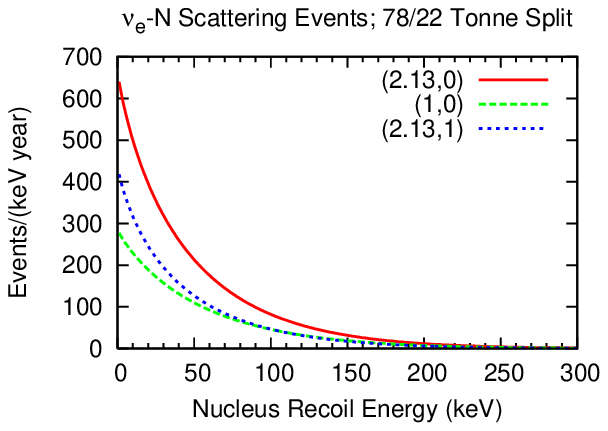}}
\scalebox{1.2}{\includegraphics{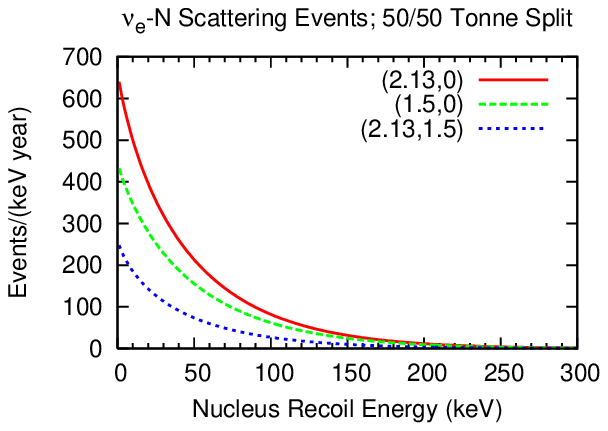}}
\caption{(Color Online) Number of neutrino-nucleus elastic scattering events per year per nuclear recoil energy in 100 tonne fiducial volume of liquid Ar. Pairs in the key denote the radii of the fiducial volume regions - the first number is the location of the outer radius and the second number is the location of the inner radius. Beam configuration is the same as for other calculations throughout the paper, with beam run at $\gamma =14$. Solid line shows calculated events in entire fiducial volume, long dashed line shows events in inner region and short dashed line shows events in outer region. Left panel is the case with boundary between inner and outer region at r=1m, inner region containing 22 tonnes Ar and outer region containing 78 tonnes. Right panel is for the case with boundary at r=1.5m, each region containing 50 tonnes of Ar.}
\label{fig:eventcuts}
\end{figure}

\begin{figure}
\scalebox{1.5}{\includegraphics{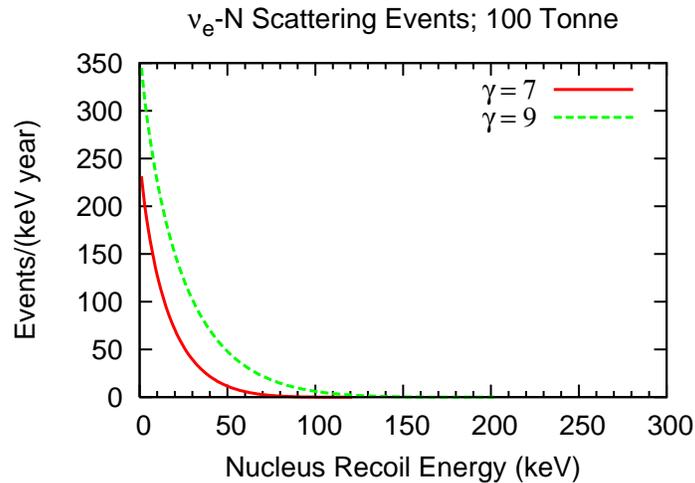}}
\caption{(Color Online) Number of neutrino-nucleus elastic scattering events per year per nuclear recoil energy in 100 tonne fiducial volume of liquid Ar. Experimental configuration is the same as for other calculations throughout the paper. Solid line shows calculated events for beam run at $\gamma = 7$, long dashed line shows events for beam run at $\gamma = 9$.}
\label{fig:events79}
\end{figure}

Returning to the cylindrical detector, we calculate the number of neutrino-nucleus elastic scattering events per year per nuclear recoil energy by folding the average flux (Equation \ref{flux}) with the differential cross section for neutrino-nucleus elastic scattering:
\beq
\frac{dN}{dT}(T) = N_t \int^{E_L^{\rm max}}_{{\rm min}(T)} \Phi(E) \frac{d\sigma}{dT}(E,T)\,dE
\eeq
Here, $E$ is the neutrino energy in the lab frame, $T$ is the nucleus recoil energy, ${\rm min}(T)=(T +\sqrt{T^2 + 2TM})/2$ is the minimum neutrino energy required to give a nucleus of mass $M$ recoil energy $T$, and $N_t$ is the number of targets in the detector. We use the cross section given in \mycite{bmr}. We neglect radiative corrections and use the analytic form factor given in \mycite{engel}. Calculated event rate curves are given in \fref{fig:eventcuts} and \fref{fig:events79}. In \fref{fig:eventcuts} we show events in the inner cylinder and outer hollow cylinder regions for a fiducial region of 100 tonnes. The boundary of the regions for the plots in the left panel of \fref{fig:eventcuts} is at r=1m; the inner and outer regions have masses of 22 and 78 tonnes, respectively. The boundary of the regions in the right panel of 
\fref{fig:eventcuts} is at r=1.5m; the inner and outer regions each have mass of 50 tonnes. For comparison, we show in \fref{fig:events79} the events occurring in the entire fiducial volume for the case of a beam run at two different boost factors, $\gamma=7,9$. 

We see from \fref{fig:eventcuts} that event rate curves will be different for events occurring in different regions of a detector. Notice that for the case of the boundary at r=1m, the left panel of \fref{fig:eventcuts}, there are more events in the outer region, while for the case of the boundary at r=1.5m, the right panel of \fref{fig:eventcuts}, there are more events in the inner region.  The average flux in the inner region is highest for the case of the boundary at r=1m, but this region has the smallest mass and hence less targets.  This is consistent with our qualitative observations in the previous section.

Recall the importance of different flux shapes for the measurements proposed in References 
\cite{mclaughlin,BJV,J&M,S&V,bueno}. To illustrate our method, suppose a measurement required a beam to be run for a year at $\gamma=13$ and then another year at $\gamma=14$.  The measurement could likely also be accomplished by running the beam at $\gamma=14$ using one detector and taking events from two separate regions for a duration of two years. We now discuss how using our method to produce different flux curves in a detector could also reduce total run time for some experiments.  Suppose a measurement required running the beam for a year at $\gamma=7$ and $\gamma=9$. Instead, the beam could be run for one year at $\gamma=14$ and events could be taken from different regions of the detector.  This situation is demonstrated by figures (\ref{fig:eventcuts}) and (\ref{fig:events79}). Each curve in \fref{fig:events79} would require a year of beam time - for a total of two years for the measurement. However, only one year is required to produce \emph{both} dashed curves in, for example, the right panel of \fref{fig:eventcuts}. The dashed curves in the right panel of \fref{fig:eventcuts} not only show more events, they show events for higher recoil energies as well. Thus we see with our method, not only can systematic uncertainties be reduced, for certain measurements more events can be obtained in less time.  

\section{Conclusions}
We have derived an expression which offers a way to quantify the fraction of neutrinos emitted from a point on the straight section that are contained in a given angle in the lab frame.  We have calculated flux shapes and events rates for different  detector configurations.  Different flux shapes can be obtained by running the beam at a fixed energy and using detectors with different dimensions, or by using different regions of a single detector.  The latter method has some advantages, including providing more events for certain measurements and reduced uncertainties. Although detailed studies of uncertainties remain to be done, we anticipate that the latter method will have less systematic uncertainties than the methods of running the beam at different energies or using different shaped detectors. 

\section{Acknowledgments}
This work was supported by the U.S. Department of Energy under Grant No. DE-FG02-02ER41216. The authors wish to thank A. B. Balantekin, J. Carlson, A. Curioni, and K. Scholberg for useful discussions.

\appendix
\section{Calculating Neutrino Flux in Lab Frame}
\subsection{Rest Frame Spectrum}
We start with the beta decay rate in the nucleus rest 
frame \cite{FFNI}:
\beq
\lambda = \int_0^{E_{\nu}^{\rm max}}
\frac{{\rm ln} 2}{m_e^5(ft)}F(\pm Z, E_e) E_e p_e E_{\nu}^2 dE_{\nu}. \label{lambda}
\eeq
Here $m_e$ is the electron mass, $(ft)$ is the ft-value containing the nuclear matrix element, 
$E_e$ and $p_e$ are the electron energy and momentum respectively, $E_{\nu}$ is the neutrino energy (we set the 
neutrino mass $m_\nu=0$), and $F(\pm Z, E_e)$ is the coulomb correction factor. The coulomb correction factor accounts for the the coulomb interaction between the emitted electron (or positron) with the charge of the nucleus.


The energy of the emitted electron depends on the energy of the emitted neutrino,
\beq
E_e = Q_{\rm nuc} - E_{\nu}, \label{eenergy}
\eeq
where $Q_{\rm nuc}$ is the nuclear mass difference (not to be confused with the atomic mass difference) 
for the transition. The electron momentum is determined from the electron energy by the usual relativistic 
expression
\beq
p_e = \sqrt{E_e^2 - m_e^2}.
\eeq
The neutrino has maximum energy when the electron has zero momentum. From \eref{eenergy} we see this is 
\beq
E_{\nu}^{\rm max} = Q_{\rm nuc} - m_e.
\eeq
The half life for the decay is
\beq
t_{1/2} = \frac{{\rm ln} 2}\lambda = \tau {\rm ln} 2, 
\eeq
where $\tau = 1/\lambda$ is the lifetime.

The integrand of \eref{lambda} is
\beq
f_R(E_\nu)=\frac{{\rm ln} 2}{m_e^5(ft)}F(\pm Z, E_e) E_e p_e E_\nu^2. \label{restspect}
\eeq We will refer to the function $f_R(E_\nu)$ as the neutrino
energy spectrum in the rest frame. Note that
\beq  \label{norm-prob}
\frac{f_R(E_\nu)dE_\nu}{{\displaystyle \int}^{\enumax}_0 f_R(E_\nu)dE_\nu} = \frac 1 \lambda f_R(E_\nu)dE_\nu 
\eeq
is the normalized probability that the neutrino will be emitted with energy in the range
$(E_\nu,E_\nu+dE_\nu)$ when a nucleus undergoes $\beta$-decay. The expressions on each side of \eref{norm-prob} are dimensionless. 

\begin{figure}
\scalebox{0.5}{\includegraphics{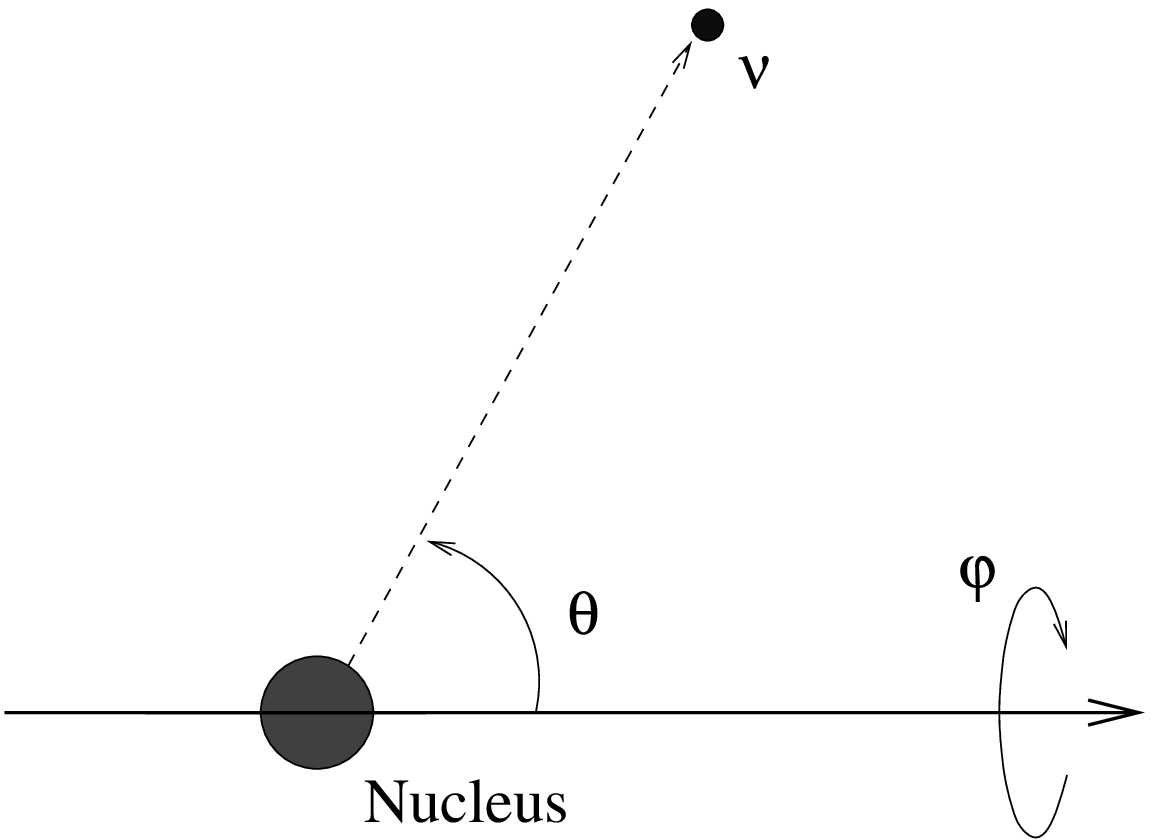}}
\caption{Coordinate system defined in the rest frame of the nucleus.}
\label{fig:rangle}
\end{figure}

There is no preferred direction for the neutrino to be emitted in the nucleus rest frame; the decay is isotropic in direction.
For the coordinate system defined in \fref{fig:rangle}, $\theta\in[0,\pi]$ and $\varphi\in[0,2\pi]$, there is an equal probability of the neutrino being emitted in the range $(\theta,\theta+d\theta)$ for any angle $\theta$.
We can define a new function for the  neutrino spectrum in the rest frame which is also a function of the direction cosine coordinate $\costh \in [-1,1]$:
\beq
f_R(E_\nu,\costh) = f_R(E_\nu). \label{theta-flux}
\eeq
This function is constant in \costh.  Note that 
\beq
\int^1_{-1} \int^{\enumax}_0 f_R(E_\nu,\costh)dE_\nu\,d\costh = 2\,\lambda. \label{theta-int}
\eeq
Now we have 
\beq
\frac{f_R(E_\nu,\costh)}{2\,\lambda}dE_\nu, d\costh \label{theta-prob}
\eeq
for the normalized dimensionless probability that the neutrino is emitted with direction cosine 
$\in (\costh,\costh+d\costh)$ and with energy $\in (E_\nu, E_\nu+dE_\nu)$.

\subsection{Transformation to Lab Frame} \label{appendix-transformation}
Now consider the rest frame moving with velocity $v$ with respect to the lab frame. Referring again to figure \fref{fig:rangle}, we take the nucleus rest frame to be moving in the direction of the arrow on the axis shown.  The neutrino emitted in the rest frame has energy $E_{ R}$ and the longitudinal and transverse components of momentum are $p_{ R}^l$ and 
$p_{ R}^t$, respectively. (We have dropped the subscript $\nu$ for clarity.) The lab frame energy and momentum components are:
\bea
E_{ L} &=& \gamma E_{ R}(1 + v \costh_R)  \label{elab}\\
p_{ L}^l &=& \gamma E_{R}(v + \costh_R) = E_L \costh_L \label{plabl}\\
p_{ L}^t &=& E_{ R} {\rm sin}\theta_R = E_L {\rm sin}\theta_L. \label{plabt}
\eea
Recall that $v = \sqrt{\gamma^2 - 1}/\gamma$. Equations \ref{plabl} and \ref{elab} can be combined to obtain the expression 
\beq
\costh_L = (\costh_R + v)/(1 + v \costh_R). \label{thetaL}
\eeq

To derive an expression for the neutrino spectrum in the lab frame, we start with \eref{theta-prob}. The transformation equations (\ref{elab}) - (\ref{plabt}) can be inverted to give the rest frame coordinates in terms of the lab frame coordinates:
\bea
E_R = E_R(E_L, \costh_L) \\
\costh_R = \costh_R (\costh_L).
\eea
Then we have 
\bea
&&\frac{f_R(E_R, \costh_R)}{2\,\lambda} dE_R\, d\costh_R \longrightarrow \nonumber \\
&&\frac{f_R[E_R(E_L, \costh_L), \costh_R(\costh_L)]}{2\,\lambda} \frac{\partial(E_R,\costh_R)}{\partial(E_L,\costh_L)}
dE_L\, d\costh_L, \label{trnsfrmtn}
\eea
where the Jacobian is ${\partial(E_R,\costh_R)}/{\partial(E_L,\costh_L)} = 1/\gamma(1 - v \costh_L)$.
Finally we have
\beq
f_L(E_L, \costh_L) dE_L\,d\costh_L = \frac{f_R[E_R(E_L,\costh_L)]}{2\,\lambda}\frac{dE_L\, d\costh_L}{\gamma(1 - v \costh_L)}.
\eeq
For an example to demonstrate this coordinate transformation we check the normalization: 
\bea
1 &=& \int^{1}_{-1}\int^{E_R^{\rm max}}_0 \frac{f_R(E_R)}{2\,\lambda} dE_R\,d\costh_R \\
  &=& \int^{1}_{-1}\int^{E_L^{\rm max}}_0 \frac{f_R[E_R(E_L, \costh_L)]}{2\,\lambda} 
                       \frac{dE_L\,d\costh_L}{\gamma(1-v \costh_L)}                   \\
  &=& \int^{1}_{-1}\int^{E_L^{\rm max}}_0 f_L(E_L, \costh_L) dE_L\,d\costh_L.
\eea
For a $\beta$-decay nucleus boosted to speed $v$ we interpret $f_L(E_L, \costh_L) dE_L\,d\costh_L$ as the normalized
probability that the neutrino will be emitted with lab energy $\in (E_L, E_L+ dE_L)$ and lab direction cosine
$\in (\costh_L, \costh_L + d\costh_L)$.  
We illustrate this with a thought experiment. Consider a large spherical neutrino detector. We send a nucleus boosted to speed $v$ through a central axis of the detector. The detector is large enough that the nucleus will decay before it reaches the end of the axis. We are able to identify the position of the nucleus when it decays and we detect the neutrino in the detector. Thus we know the lab direction cosine that the neutrino was emitted with. We also measure the neutrino's energy when we detect it. Now we repeat the experiment with another nucleus over and over. The distribution of the ``measured'' neutrino energy and direction cosines will be given by $f_L(E_L, \costh_L)$.

\subsection{Number of $\beta$-decays per second in the Ring}
Consider the storage ring.  Initially, at time $t=0$, there are zero nuclei in the ring. We start injecting $\omega$ ions per second into the ring at $t=0$. We would like to know the total number of decays per second that occur in the ring after an equilibrium situation has been reached.  In the lab frame, the decay rate is $\lambda_L = 1/\tau_L$, where $\tau_L$ is obtained from the lifetime in the rest frame by $\tau_L = \gamma\tau_R$.  Let $N(t)$ be the number of nuclei in the ring at time $t$, and $dN$ be the change in the number of nuclei during time interval $dt$. Then we have
\beq
dN = - N \lambda_L dt + \omega dt.
\eeq
Using the initial conditions, the solution to this differential equation is
\beq
N(t) = \frac \omega {\lambda_L} - \frac{\omega}{\lambda_L}e^{-\lambda_L t}.
\eeq
Equilibrium is reached in the limit $t\rightarrow \infty$. The number of \emph{nuclei} in the ring, $N(t)$, after equilibrium is reached is 
\beq
\lim_{t \rightarrow \infty} N(t) = \frac \omega {\lambda_L} = \gamma\tau_R\,\omega.
\eeq
The number of \emph{decays} in the ring, $\lambda_L\,N(t)$, after equilibrium is reached is
\beq
\lim_{t \rightarrow \infty} \lambda_L N(t) = \omega.
\eeq
After a long time of injecting ions into the ring, the number of $\beta$-decays occurring in the entire ring is equal to the number of ions per second injected into the ring.

\subsection{Integrating Over the Ring and Detector} \label{appendix-integrating}
A cylindrical detector, with radius $r$ and length $h$, is placed at a distance $d$ away from the ring along the axis of one of the straight sections.  The flux at each point in the detector is not the same since neutrinos are incident from different angles. To compute the number of events in a detector we need to fold the flux with the cross section for the neutrino scattering interaction. We could calculate the flux at each point in the detector, multiply the flux at each point by the cross section, integrate over position in the detector, and then integrate over neutrino energy. However, we are not interested in any angular dependence of events. Rather than calculating a flux at each point in the detector, we will use the average of the flux at all points. To get the total events we take the average flux at a point, multiply by the number of targets in the detector (i.e. the number of points), and fold this with the cross section. For the purposes of calculating total events this is equivalent to the preceding way to do the calculation.  However, one should keep in mind that if angular dependence of the events is desired, perhaps for considering detection mechanisms, the former method is necessary.

We first consider the case of an infinitesimally thin detector. To calculate the average flux in this detector we add the total number of neutrinos per second passing through the slice and then divide by the area of the slice. Let 
$N_\nu(E_L)\, dE_L$ be the total number of neutrinos per second, with lab energy $\in (E_L, E_L + dE_L)$, passing through detector. 

The $\beta$-decays occur throughout the entire storage ring. For a ring of total length $L$, with straight section lengths $S$, the number of decays per second per length element $dx$ occurring at any point along the ring is $\omega/L$. To find 
$N_\nu(E_L)\, dE_L$ we sum the contribution of neutrinos from points along the straight section, 
$(\omega/L)\,f_L(E_L, \costh_L) dE_L\,d\costh_L$, that reach the detector:
\beq
N_\nu(E_L)\, dE_L = \omega\int^1_{\frac{x+d}{\sqrt{(x+d)^2 + r^2}}} \int^S_0 f_L(E_L, \costh_L) 
\frac{dx}{L}\,d\costh_L\,dE_L.  \label{Ndisk}
\eeq 
For a detector with thickness $h$, \eref{Ndisk} (with $x+d \rightarrow x+d+z$) is the number of neutrinos per second passing through the slice of the detector located between $(z, z+dz)$. Integrating over $dz/h$ gives the average of these numbers over all slices. Finally, the average of the flux over all points in the detector is obtained by dividing this average over slices by the cross sectional area of the detector: 
\beq
\Phi(E_L) = \frac{\omega}{\pi r^2}\int^1_{\frac{x+z+d}{\sqrt{(x+z+d)^2 + r^2}}} \int^S_0 \int^h_0 f_L(E_L, \costh_L) 
\frac{dz}{h}\,\frac{dx}{L}\,d\costh_L. \label{flux}
\eeq 
Notice that \eref{flux} has dimensions of number per energy per time per area.

\newpage

\end{document}